\documentstyle[12pt,bezier]{article}

              \def\be{\begin{equation}}
              \def\ee{\end{equation}}
              \def\bea{\begin{eqnarray}}
              \def\eea{\end{eqnarray}}
        \def\ba{$$\begin{array}}
        \def\ea{\end{array}$$}
\begin{document}
\begin{center}
{\Large{\bf{ A Modified Quantum Renormalization Group  \\
for  \\
{\bf xxz} Spin Chain}}}
\vskip 0.2 cm   
{{{\bf A. Langari}\footnote{E-mail : \it {langari@netware2.ipm.ac.ir}} }} , and {\bf V. Karimipour}
\vskip .2cm
{\it \small { Institute for Studies in Theoretical Physics 
and Mathematics
\\ P.O.Box: 19395-5531, Tehran, Iran. \\  Department of Physics, 
Sharif University of Technology\\ P.O.Box:11365-9161, Tehran, Iran  }}
\end{center}
\vskip .1cm
\begin{abstract}
A simple modification of the standard Renormalization Group (RG) technique
for the study of quantum spin systems is introduced.
Our method which takes into account the effect of boundary conditions
by employing the concept of superblock, may be regarded as a simple
way for obtaining first estimates of many properties of spin systems.
By applying this method
to the {\bf xxz} spin $\frac{1}{2}$ Heisenberg chain, we obtain the ground state 
energy with much higher accuracy than the standard RG.
We have also obtained the staggered magnetization and the z-component of  
spin-spin correlation function which confirms the absence of long-range order
in the massless region of the 1D {\bf xxz} model.
\end{abstract} 

PACS numbers : 05.50.+q,75.10.-b,75.10.Jm \\
Keywords : Quantum renormalization group, Spin systems, Heisenberg model.

\newpage
\def\bra#1{{\langle #1 \vert}}
\def\ket#1{{\vert #1 \rangle}}
\def\x#1{{\sigma_{#1}^{x}}}
\def\z#1{{\sigma_{#1}^{z}}}
\def\y#1{{\sigma_{#1}^{y}}}
\def\d{\Delta}
\def\a{\alpha}
\def\b{\beta}
\def\ge{\ground state energy}
\def\ep{\varepsilon}

\section{Introduction}  

Soon after the introduction of real space Renormalization Group (RG)
by Wilson and it's application to the Kondo problem$^1$, it was found
that this type of RG does not always produce accurate numerical
results, as compared with other powerful techniques.
One of the weak performances of the
RG method was found in Lee's work on 2D Anderson localization$^2$.
Using numerical RG method he concluded that there is a critical amplitude
for the random potential which induces localization. However, not long after
Lee's work, Lee and Fisher$^2$, using a different approach,
found that the 2D model is logarithmically
localized even for arbitrary small randomness. This
result is now generally accepted.
Such unsatisfactory behavior  of the RG method compared with the 
other methods such as Quantum Monte Carlo,
was one of the reasons that the RG method for quantum lattice problems
remained undeveloped during
the 80's.

Let us first describe the standard RG method$^3$ ( see ref.4 for a modern treatment and references therein).
First, one divides
the whole lattice into small blocks (fig.1) and obtains the lowest
states ($\ket{i}$) of each isolated block for a particular
Boundary Condition (BC).The effect of inter-block
interactions is taken into account
by constructing an effective Hamiltonian $ H^{eff} $ which now acts on a smaller
Hilbert space $ {\cal H }^{eff} $ embedded in the original one. 
In this new Hilbert space, each 
of the former blocks is treated as a single site . \\
The technical way of implementing this idea is to construct an embedding
operator $$ T : {\cal H}^{eff} \longrightarrow {\cal H } $$  and 
a truncation operator 
$$ T^{\dagger} : {\cal H} \longrightarrow {\cal H }^{eff} $$
and demand the commutativity of the following diagram$^4$ : \\
$$
\begin{array}{c}
{\cal H}^{eff} \stackrel{T}{\longrightarrow}  {\cal H } \\
H^{eff}  \downarrow  \hspace{1.5cm} \downarrow \hspace{0.5cm} H \\
{\cal H}^{eff} \stackrel{T}{\longrightarrow}  {\cal H } \\
\end{array}
$$
i.e. : $T H^{eff} = H T $.
From the last relation one obtains the effective Hamiltonian as :
\be H ^{eff} = T^{\dagger} H T \ee
Note that the operators $ T $  and $ T^{\dagger } $ satisfy the 
relation $ T^{\dagger} T = 1_{{\cal H}^{eff}} $ 
but $ T T^{\dagger } \ne 1_{{\cal H }}. $ 
More precisely one divides $H$ as $H=H^{B}+H^{BB}$ (fig.1) where  
$H^{B}=\sum_{I}h_{I}^{B}$ is the sum of block Hamiltonians
and $H^{BB}=\sum_{I,J}h_{I,J}^{BB}$ is the sum of inter-block Hamiltonians
and constructs $T=\prod_{I}T_{I}$ where $T_{I}$ is constructed as follows :
\be
T_{I}=\sum_{i=1}^{m}\ket{i}\bra{i}
\ee
where m is the number of low energy states kept.
Note that each $T_{I}$ acts trivially on all the other blocks $J \ne I$
consequently $[T_{I},T_{J}]=0$.
Repeating the RG steps one hopes to restrict himself 
to spaces of lower and lower energy
and finally arrives at the ground state energy.
In the general case this process induces a renormalization of coupling constants :
\be
H^{eff} \equiv H(k'_1,k'_2, ...) = T^{\dagger} H(k_1,k_2, ...) T
\ee
where $K = (k_1,k_2, ...)$ is the coupling constant space.

In principle the form the Hamiltonian changes in a such RG program rendering 
analytical treatment, rather difficult if not possible. In the standard RG
treatment of some models including ${\bf xxz}$ Heisenberg chain ,
it is possible to choose the embedding operator so that no new terms appear
in the renormalized Hamiltonian$^3$. In such cases, repeating the RG steps, 
leads to the following general form : 
\be
T^{\dagger ^n} H(K) T^n = \sum_{p=0}^{n-1}\frac{N}{l^{p+1}}e_{o}^{B}(K^{(p)}) 
+ H_{\frac{N}{l^n}}(K^{(n)})
\ee
here $H_{N} $ is the Hamiltonian of the N-sites lattice, $l$ is the number of
sites in each block, $n$ is the number of RG steps performed, 
$e_{o}^{B}(K^{(p)})$ is the ground state energy of each block 
after the p-th step and under
RG : $ K^{(n)} \longrightarrow K^{(n+1)} $,  
then the ground state 
of the whole lattice is then given by : 
\be
E_o = \sum_{p=0}^{\infty}\frac{N}{l^{p+1}}e_{o}^{B}(K^{(p)})
\ee
           
However the main difficulty of the method is that by 
fixing a particular
BC on an isolated block one may loose a number of states which contribute to the 
ground state of the whole lattice due to the interaction of the block
with its surrounding. 
Stated in another way, the ground state wave function of the whole 
lattice for a particular BC is not a simple juxtaposition of the ground 
state wave functions of the blocks.
This point is clearly
highlighted in 1D Tight Binding model in which the standard RG
fails for some type of BCs$^{5,6}$. It must be noticed that
this difficulty is not removed by increasing the size of blocks.

In the Density Matrix Renormalization Group$^7$ one embeds each 
block into a larger(super) block and considers the block as a quantum
system in interaction with a reservoir(the rest of the superblock). 
The block can then be described by
a reduced density matrix, whose eigenkets with the highest eigenvalues
are used to construct the embedding operator($T$).
Provided that one keeps a large number of states in each step
of truncation(i.e. $ m = 16, 24, 36 $ and $ 44 $ in four different runs$^7$),
this method leads to 
very accurate results for energies and correlation functions of
the isotropic {\bf xxx} Heisenberg chain.

However keeping a large number of states renders the problem of identification
of coupling constants of the effective Hamiltonian very difficult, which means
that one should make a compromise in obtaining accurate results for energies and
correlation functions on the one hand and coupling constant trasformations
on the other, the later objective is of course important if one is interested
in studying the critical properties of these systems. In fact while the DMRG
method is quite appropriate for obtaining the former properties, the standard
RG appeares to give better results for the later properties$^8$. One should
adapt one or another of these RG schemes to obtain reasonable results in both
directions. In ref.[9] a strategy has been suggested for the adaptation
of DMRG so that one can also derive the transformation of coupling constants
and the critical properties of the {\bf xxz} Heisenberg chain.

The present paper has a different starting point in that it adapts the
standard RG and incorporates in it the interaction of the block with its
environment, to do the same general task. The method presented in this paper
and in ref.[9] are to be regarded as complementary. Depending on the problem
and the desired accuracy either of them can be used.
The simplicity of our method is such that one may hope
to apply it to two dimensional systems.
Although it does not yield as
much accuracy as in DMRG, is much simpler practically, so that for obtaining 
first estimates of many properties of spin or fermion systems one may try
this method$^{10}$.
For the ${\bf xxz}$ chain,
with blocks of 3 sites in the infinite chain and with
retaining only $m=2$ states in each block, we have obtained the ground state
energy with an accuracy much better than the standard RG method. The    
results are compared  in table-1 and table-2.
We have also obtained the staggered magnetization and spin-spin correlation
functions in the massless region of the {\bf xxz} model. These results
verify the absence of long-range order in this region which is in agreement
with the known results. 

\section{A modified RG for the $s=\frac{1}{2}$ {\bf xxz} chain }

The system under
consideration is the anti-ferromagnetic $s=\frac{1}{2}$ {\bf xxz} chain with the
following Hamiltonian :
\be
H_{N}(J,\Delta)=(J/4) \sum_{i=1}^{N-1}(\x{i}\x{i+1}+\y{i}\y{i+1}+
\Delta(\z{i}\z{i+1}))
\ee 
where $J$ is positive and $\d$ is the anisotropy parameter
which is limited by $(0\leq\Delta\leq1)$ in the massless region.

Fig.2 shows the decomposition of the chain into isolated blocks and super blocks
with an odd number of sites per each block.
Thus the block Hamiltonian $h^{B}$ and the superblock Hamiltonian $h^{SB}$ are 
\be
h^{B}=(J/4)(\x{2}\x{3}+\x{3}\x{4}+\y{2}\y{3}+\y{3}\y{4}+
\Delta(\z{2}\z{3}+\z{3}\z{4}))
\ee
\be
h^{SB}=(J/4)(\x{1}\x{2}+\y{1}\y{2}+\Delta\z{1}\z{2})+h^{B}+
       (J/4)(\x{4}\x{5}+\y{4}\y{5}+\Delta\z{4}\z{5})
\ee
The basic steps of our method are follows : 

{\bf step-1)} First one diagonalizes the superblock Hamiltonian ($h^{SB}$).
Although $h^{SB}$ does not have su(2) symmetry, it has a number of symmetries  
which helps in it's diagonalization. Defining the z-component of total spin
$S^{z}=\frac{1}{2}(\z{1}+\z{2}+\z{3}+\z{4}+\z{5})$, the flip operator
$\sigma=\x{1}\x{2}\x{3}\x{4}\x{5}$, and the parity operator($\Pi$),
$\Pi\ket{\alpha_{1}\alpha_{2}\alpha_{3}\alpha_{4}\alpha_{5}}=
\ket{\alpha_{5}\alpha_{4}\alpha_{3}\alpha_{2}\alpha_{1}}. $    
It is easy to check that, $H$ commutes with 
$\Pi ,\sigma $ and $S^{z}$. Furthermore 
$\left[ \Pi , S^{z} \right]=0,
\left[ \Pi , \sigma \right]=0,$
which means that the energy eigenstates of $h^{SB}$ can be chosen 
to have definite parity and the z-component of spin. 
The ground state is doubly degenerate, one with the z-component of
the spin equal to $+\frac{1}{2}$ and the other equal to $-\frac{1}{2}$.
\be
h^{SB}\ket{\pm\frac{1}{2}}_{SB}=e_{o}^{SB}(J,\d)\ket{\pm\frac{1}{2}}_{SB}
\ee  
where $\ket{\pm\frac{1}{2}}_{SB}$ are the doubly degenerate ground state
of $h^{SB}$ and $e_{o}^{SB}$ is their corresponding energy.
We note that the states of lowest energy are 
even parity states with their z-component of spin equal to $\pm\frac{1}{2}$.
Thus the generic form of $\ket{+\frac{1}{2}}_{SB}$ is as follows.
\bea
\ket{+\frac{1}{2}}_{SB}&=&a_{1}\ket{+++--}+a_{2}\ket{++--+}+a_{3}\ket{++-+-}
\nonumber \\
&+&a_{4}\ket{+-++-}+a_{5}\ket{+-+-+}+a_{6}\ket{-+++-}+a_{4}\ket{-++-+}
\nonumber  \\
&+&a_{3}\ket{-+-++}+a_{2}\ket{+--++}+a_{1}\ket{--+++}
\eea  
The state $\ket{-\frac{1}{2}}_{SB}$ is obtained from $\ket{+\frac{1}{2}}_{SB}$
by flipping all the spins. All the coefficient $a_{i}$ ; $i=1,\ldots,6 $
can be expressed as functions of $J$ and $\d$ 
although we do not need to explicitly write them at this step, 
and $e_{o}^{SB}$,
itself can be calculated numerically to any desired accuracy.

{\bf step-2)} Knowing that the ground state of the block is a 2-dimensional
subspace with $s_{z}=\pm\frac{1}{2}$ ,
we then project the $\ket{\pm\frac{1}{2}}_{SB}$ states onto 
the $s_{z}=\pm\frac{1}{2}$ 
subspaces of the block Hilbert space. It is this step 
which effectively smoothes 
out the sharp effects of the boundary conditions,
by immersing the block into a superblock (which to some extent simulates 
the effect of the rest of the lattice),  
the projected state does not restrict to any
particular boundary conditions of an isolated block (open, periodic, ...).
Moreover, in this way the form of the Hamiltonian does not change, as we will see
later. What we do is that, 
after projection onto
the block Hilbert space we implement the projection operator $P^{+}$ which is 
shown below, to reach the $s_{z}=+ \frac{1}{2}$  subspaces of the block Hilbert space.
\be
P^{+} = \ket{++-}\bra{-++} + \ket{+-+}\bra{+-+} + \ket{-++}\bra{++-}  \nonumber
\ee

Then the resulting normalized states are 
\bea
\ket{+\frac{1}{2}}_{B}\equiv P^{+} \ket{+\frac{1}{2}}_{SB} =
(1/A)((a_1+a_4)\ket{++-}+2a_3\ket{+-+}+(a_1+a_4)\ket{-++})   \cr
\ket{-\frac{1}{2}}_{B}\equiv P^{-} \ket{-\frac{1}{2}}_{SB} =
(1/A)((a_1+a_4)\ket{--+}+2a_3\ket{-+-}+(a_1+a_4)\ket{+--})
\eea   
where $A$ is a normalization factor.
Straightforward calculations give the following results 
\bea
&\a \equiv a_1 + a_4=(\frac{2\ep^{2}+\ep\d-2}{2\ep(\ep^{2}-\d^{2}-1)})+\frac{\ep}{2}&  \nonumber \\
&\b \equiv 2 a_3= \frac{(\ep-\d)(2\ep^{2}+\ep\d-2)}{\ep(\ep^{2}-\d^{2}-1)}&        
\eea
where    
\be
\ep=(2/J)e_{o}^{SB}.
\ee  
Numerical calculation gives the following result for $e_{o}^{SB}$ $(0\leq\Delta\leq1)$,
\be
e_{o}^{SB}(J,\d)=J(a+b\d+c\d^{2}+d\d^{3}+e\d^{4}+f\d^{5}+g\d^{6}+h\d^{7}
+k\d^{8}+l\d^{9})
\ee    
with 
$$
\begin{array}{lllll}
a=-1.36623   &  b=-0.63707  &   c=2.92264   &   d=-22.07630 &   e=84.52129 \\
f=-187.92379 &  g=251.64014  &   h=-200.06805 &  k=86.98911  &    l=-15.92981 .\\
\end{array}
$$

{\bf step-3)} The embedding operator is now constructed  
in the following form 

\be
T_{I}=(\ket{+\frac{1}{2}}_{B}\bra{+})+(\ket{-\frac{1}{2}}_{B}\bra{-})
\ee \\
where $\ket{+}$ and $\ket{-}$ are the renamed base kets of the 
effective Hamiltonian Hilbert space of each block.

Having the form of the embedding operator, one can calculate the
projection of any operator onto the effective Hilbert space, as in eq.(1) i.e. 
\bea
&\sigma_{i}^{\mu,eff}=T^{\dag}\sigma_{i}^{\mu}T=(\frac{2\a\b}{A^{2}})\sigma_{i}^{'\mu}
\hspace{2cm}i=1,3 ;\mu=x,y&  \nonumber \\ 
&\sigma_{i}^{z,eff}=T^{\dag}\sigma_{i}^{z}T=(\frac{\b^{2}}{A^{2}})\sigma_{i}^{'z}
\hspace{2cm}i=1,3&
\eea
where $\sigma'_{i}$ acts as pauli matrices in the new Hilbert space.
Fig.1 shows that the interaction between blocks are in the following form
\be
h_{I,J}^{BB}=(J/4)(\x{3I}\x{1J}+\y{3I}\y{1J}+\d\z{3I}\z{1J}) .
\ee  
Then the effective Hamiltonian between the new sites are 
\be
h_{I,J}^{eff}=T_{J}^{\dag}T_{I}^{\dag}h_{I,J}^{BB}T_{I}T_{J}
\ee    
where I and J are two neighbouring blocks while $T_{I}$ and $T_{J}$ are
their corresponding embedding operators. Doing so we arrive at :
\be
h_{I,J}^{eff}=(J'/4)(\sigma_{I}^{'x}\sigma_{J}^{'x}+\sigma_{I}^{'y}\sigma_{J}^{'y}
+\d'\sigma_{I}^{'z}\sigma_{J}^{'z})
\ee  
where
\bea
J'=(\frac{2\a\b}{A^{2}})^{2}J   \cr
\d'=(\frac{\b}{2\a})^{2}\d
\eea 
Eq.(21) describes the RG flow of J and $\d$. The
RG flow goes to the origin of $J-\d$ plane which verifies that we are
in the massless region.

{\bf step-4)} The crudest estimate for the ground state energy is just the sum of 
ground state energies of all the blocks or superblocks. Thus 
\be
(\frac{E}{N})^{(0)}=\frac{1}{5}e_{o}^{SB}(J,\d)
\ee
or
\be
(\frac{E}{N})^{(0)}=\frac{1}{3}e_{o}^{B}(J,\d) 
\ee 
where $(\frac{E}{N})^{(0)}$ is the ground state energy of the whole system
in the zeroth order approximation. Eq.(23) in fact defines 
$e_{o}^{B}$ as :
\be 
e_{o}^{B}=\frac{3}{5}e_{o}^{SB}
\ee  

However this estimate clearly neglects the contribution to the energy from 
interaction between blocks. To take this missing contribution into account,
we add to $(\frac{E}{N})^{(0)}$, the ground state energy of the new lattice, and obtain :
\be
(\frac{E}{N})^{(1)}=\frac{1}{3}e_{o}^{B}(J,\d)+\frac{1}{9}e_{o}^{B}(J^{(1)},\d^{(1)})
\ee 
iterating this procedure we finally obtain 
\be
\frac{E}{N}=\sum_{n=0}^{\infty}\frac{1}{3^{n+1}}e_{o}^{B}(J^{(n)},\d^{(n)})
\ee
or                                          
\be
\frac{E}{N}=\sum_{n=0}^{\infty}\frac{1}{3^{n+1}}(\frac{3}{5}e_{o}^{SB}(J^{(n)},\d^{(n)}))
\ee 
where $J^{(0)}=J$ , $\d^{(0)}=\d$ and $J^{(n)}$ , $\d^{(n)}$ are the renormalized
coupling constants after n-steps of RG.
In fact the above argument describes the physical interpretation of the formal 
limiting process defined in ref.4, in which the ground state energy is obtained as : 
$$ E=lim _{n\rightarrow {\infty} } {T^{\dagger} }^n H T^{n} $$
However in standard RG the terms $ e_0^B (J, \Delta) $ is the result of 
diagonalization of $ h^B $ , i.e: $T^{\dagger} h^B T =  e_0^B $ , while in our
method $e_0^B $ is just another expression for $ \frac {l}{l'}e_0^{SB} $ where 
$e_{0}^{SB}$ is the eigenvalue of $ h ^{SB} $, $l'$ and $l$ are the sizes of the 
superblocks and blocks respectively.
The results for the ground state energy per site, obtained by this modified RG method
are collected in table-1, for a series of anisotropy parameter
$(0\leq\Delta\leq1)$. The results are compared both with the exact results 
(obtained by the Bethe ansatz for {\bf xxx}$^{11}$ and
{\bf xxz}$^{12}$ spin $\frac{1}{2}$
chains) and with those obtained from the standard RG$^{13}$. It is clearly 
seen that this type of modified RG method yields much better results than the 
standard RG method.

A remark is in order here, concerning the non-variational nature of our
results, which is caused by our crude estimate for the ground state energy
of each block in terms of that the superblock ($e_{o}^{B}=\frac{3}{5}e_{o}^{SB}$).
However one should note that this behaviour is not unusual in RG treatment
of lattice systems. For example in standard RG at second order perturbation
of free-fermion model$^3$ and isotropic Heisenberg model$^{14}$, 
one obtains non-variational results. We have also
obtained the ground state energy in the $\d=0$ ({\bf xx} model) case
for other $\frac{n_{B}}{n_{SB}}$ ratios, namely for the ratios $\frac{5}{7}$,
$\frac{7}{9}$, $\frac{9}{11}$ and $\frac{11}{13}$,
the results are collected in table-2.
It is interesting to note that our results in these cases
are above the exact values of ground state energy, which indicates that
the non-variational result in the $\frac{3}{5}$ case is due to the very small
block and superblock sizes. Therfore we expect that for $n_B$ and $n_{SB}$
not very small, our method gives indeed variational results. Our data in 
table-2 show much accuracy compared with the standard RG results for any 
value of block sizes. These results show that the computed ground state
energy indeed converges to the exact result faster than the standard RG
method by increasing the $\frac{n_{B}}{n_{SB}}$ ratios.
 
\section{Correlation Functions}
In this section we want to gain more insight on the effective Hamiltonian
constructed by this modified RG. The {\bf xxz} Hamiltonian in the massless
region $(0\leq\Delta\leq1)$ has a non-degenerate ground state with a zero
staggered magnetization in the thermodynamic limit $N {\longrightarrow} \infty$.
The staggered magnetization is defined as
\be
m_{st} = \bra{0} \frac{1}{2N} \sum_{i=1}^{N}(-1)^i \z{i} \ket{0}
\ee
In the RG formalizm the ground state $\ket{0}$ is replaced by $T \ket{0'}$,
where $\ket{0'}$ is the ground state of the effective Hamiltonian Hilbert
space. Here we consider a chain of length $3^p$ and let $p$ go to infinity.
Then the staggered magnetization can be written in the following form
\be
m_{st} =  \frac{1}{2N} \sum_{i=1}^{N}(-1)^i \bra{0'} T^{\dagger} \z{i} T \ket{0'}
\ee
By dividing the chain into 3-sites blocks and consider $T$ as the product of
the embedding operator of each block, we can use eq (17) and write it again as

$$ T^{\dagger} \z{i} T=\gamma_i \z{i}'   $$
where
\be
\gamma_1=\gamma_3=(\frac{\b^{2}}{A^{2}}) \hspace{1cm} ;\hspace{1cm}
\gamma_2=(\frac{2 \a^2 - \b^2}{A^2}).
\ee
Here we can continue this relation and replace $\ket{0'}=T \ket{0''}$, and so on.
Then the staggered magnetization will be
\be
m_{st} =  \frac{1}{2N} \sum_{i=1}^{N}(-1)^i \bra{0^{(p)}} \gamma_i^{p}
\z{i}^{(p)}  \ket{0^{(p)}}
\ee
which gives $m_{st} = 0, $ when $p$ goes to infinity, since $(0\leq \gamma_i <1)$
for $(0\leq\Delta\leq1)$.

Another quantity which verifies the absence of long-range order in the massless
region of the {\bf xxz} model is the z-component of spin-spin correlation
functions. It is defined as
\be
g(r)= \bra{0} (-1)^r \z{i} \z{i+r} \ket{0}
\ee
for the anti-ferromagnetic ordering. In the classical Neel order, $g(r)$ is
a constant for any distance $r$, but quantum fluctuations can destroy Neel
ordering. Despite the 2D Heisenberg model in the 1D {\bf xxz} model quantum
fluctuations destroy the long-range order.

We show that the constructed effective Hamiltonian will produce the above
results. Again we assume a $3^p$ sites chain and let
$p {\longrightarrow} \infty$. In this procedure $s_i^{z}$ is considered
at the center of chain and the other blocks are expanded in both sides of it
symmetrically. To calculate $g(r)$, the ground state 
$\ket{0}$ is replaced by $T \ket{0'}$ and eq.(30) is used for the effective
operator in the effective Hilbert space. The obtained $g(r)$ is plotted
in fig.3a for {\bf xx} model($\d=0$) and compared with the exact results$^{15}$. 
We have also plotted $g(r)$ for $\d=0.5$ and $\d=1$ in fig.3b.
All of these
graphs show good agreement with the exact results$^{15}$. 
The good agreement of the correlation
functions verify that the constructed effective Hamiltonian is a good
approximate one for the low energy spectrum of the {\bf xxz} model.

\section{Conclusions}

We have shown how by a simple modification of the standard RG method 
one can take care of the effect of boundary conditions in treating 
quantum lattice systems.
We have applied this method to the spin $\frac{1}{2}$ {\bf xxz} model
in the massless region. We have shown our method give more accurate
results than the standard RG for different value of block sizes.
Our data of the ground state energy converge to the exact one
faster than the standard RG results
by increasing the block sizes. We have also calculated the staggered
magnetizaton and the z-component of spin-spin correlation function in the
massless region. These data confirm that we have obtained a good approximate
Hamiltonian for the low energy spectrum which leads to no long-range order
where, $0\leq\Delta\leq1$. These results are in good agreement with the
exact results$^{11,12,15}$.
Our method although not as exact  as the more
sophisticated DMRG method has the merit that it can be implemented 
very easily on personal computers. One may hope that for a first 
estimate of many properties of these systems, our method or it's improvements
(or adaptation for other lattice systems) may give acceptable results. 
This method can also be applied to other spin systems, to fermion models$^{10}$,
and more interestingly to 2-dimensional models (work on 2D anti-ferromagnetic
Heisenberg model is in progress).

\section{Acknowledgement}
A. L.  would like to express his deep gratitude to Miguel A. Martin-Delgado
for valuable comments. We would also like to thank
M.R.H.Khajepour for an interesting discussion, and 
B.Davoodi, M.R.Ejtehadi 
and J.Davoodi for their useful comments in numerical computations.

\newpage
{\Large \bf Tables}\\

Table-1. Ground state energy per site for standard RG, exact results 
and modified RG for {\bf xxz} spin=$\frac{1}{2}$ chain.
$$
\begin{array}{|c|c|c|c|}\hline
\d&(\frac{E}{JN})_{standard RG}&(\frac{E}{JN})_{exact}&(\frac{E}{JN})_{modified RG}\\ \hline
0&-0.28284295 & -0.31830988 & -0.32529130 \\ \hline
0.1&-0.29208390 & -0.32869220 & -0.33583683   \\ \hline
0.2&-0.30163311 & -0.33956430 & -0.34603469   \\ \hline
0.3&-0.31150403 & -0.35091277 & -0.35700253   \\ \hline
0.4&-0.32171198 & -0.36272723 & -0.36813752   \\ \hline
0.5&-0.33227452 & -0.37500000 & -0.37947617   \\ \hline
0.6&-0.34321210 & -0.38772591 & -0.39120956   \\ \hline
0.7&-0.35454900 & -0.40090224 & -0.40325429   \\ \hline
0.8&-0.36631503 & -0.41452883 & -0.41546686   \\ \hline
0.9&-0.37854863 & -0.42860840 & -0.42793815   \\ \hline
1&-0.39130452 & -0.44314718 & -0.44076268     \\ \hline
\end{array}
$$

Table-2. Ground state energy per site of standard RG, exact results 
and modified RG for $\d=0$ at different value of block sizes.
$$
\begin{array}{|c|c|c|c|}\hline
Exact \,\,result & Standard \,\,RG & Modified \,\,RG & Type \,\,of \,\,SB \,\,and \,\,B\\ \hline
-0.3183 (\frac{-1}{\pi}) & n_{B}=3, -0.2828 & -0.3253 & n_{SB}=5, n_{B}=3 \\ \hline
               & n_{B}=5, -0.2927 & -0.3113 & n_{SB}=7, n_{B}=5 \\ \hline
               & n_{B}=7, -0.2983 & -0.3096 & n_{SB}=9, n_{B}=7 \\ \hline
               & n_{B}=9, -0.3019 & -0.3098 & n_{SB}=11, n_{B}=9 \\ \hline
               & n_{B}=11, -0.3044 & -0.3103 & n_{SB}=13, n_{B}=11 \\ \hline       
\end{array}
$$

\newpage
{\Large \bf Figure Captions :}\\

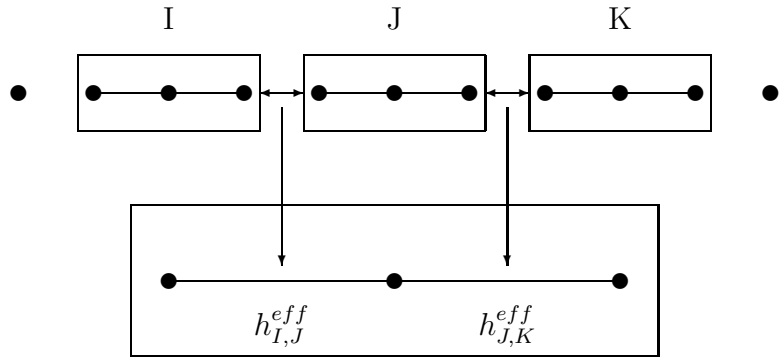
{\bf Fig.1)} Decomposition of the lattice to isolated blocks and considering 
neglected bonds as the effective interactions in the new Hilbert space.

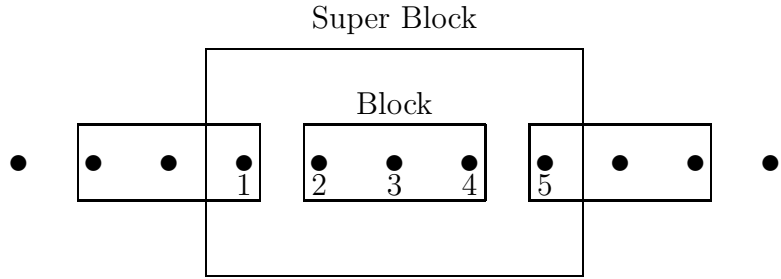
{\bf Fig.2)} Block and Super Block for the lattice chain.

{\bf Fig.3a)} z-component of spin-spin correlation function
$g(r) = \langle (-1)^r s_{i}^z s_{i+r}^z \rangle$ versus $r$ for 
{\bf xx} model($\d=0$), modified RG and exact results.

{\bf Fig.3b)} z-component of spin-spin correlation function
$g(r) = \langle (-1)^r s_{i}^z s_{i+r}^z \rangle$ versus $r$ for 
anisotropy parameter ($\d$) equal to $0.5$ and $1$, obtained by modified RG.

\newpage
\unitlength=1.00mm
\special{em:linewidth 0.4pt}
\linethickness{0.4pt}
\begin{center}  
\begin{picture}(151.00,100.00)
\put(30.00,80.00){\circle*{2.00}}
\put(40.00,80.00){\circle*{2.00}}
\put(50.00,80.00){\circle*{2.00}}
\put(60.00,80.00){\circle*{2.00}}
\put(70.00,80.00){\circle*{2.00}}
\put(80.00,80.00){\circle*{2.00}}
\put(90.00,80.00){\circle*{2.00}}
\put(100.00,80.00){\circle*{2.00}}
\put(110.00,80.00){\circle*{2.00}}
\put(120.00,80.00){\circle*{2.00}}
\put(130.00,80.00){\circle*{2.00}}
\put(38.00,75.00){\framebox(24.00,10.00)[cc]{}}
\put(68.00,75.00){\framebox(24.00,10.00)[cc]{}}
\put(98.00,75.00){\framebox(24.00,10.00)[cc]{}}
\put(62.00,80.00){\vector(1,0){6.00}}
\put(92.00,80.00){\vector(1,0){6.00}}
\put(97.00,80.00){\vector(-1,0){5.00}}
\put(67.00,80.00){\vector(-1,0){5.00}}
\put(40.00,80.00){\line(1,0){20.00}}
\put(70.00,80.00){\line(1,0){20.00}}
\put(100.00,80.00){\line(1,0){20.00}}
\put(50.00,55.00){\circle*{2.00}}
\put(80.00,55.00){\circle*{2.00}}
\put(110.00,55.00){\circle*{2.00}}
\put(45.00,45.00){\framebox(70.00,20.00)[cc]{}}
\put(50.00,55.00){\line(1,0){60.00}}
\put(65.00,78.00){\vector(0,-1){21.00}}
\put(95.00,78.00){\vector(0,-1){21.00}}
\put(65.00,49.00){\makebox(0,0)[cc]{$h_{I,J}^{eff}$}}
\put(95.00,49.00){\makebox(0,0)[cc]{$h_{J,K}^{eff}$}}
\put(50.00,90.00){\makebox(0,0)[cc]{I}}
\put(80.00,90.00){\makebox(0,0)[cc]{J}}
\put(110.00,90.00){\makebox(0,0)[cc]{K}}
\put(80.00,25.00){\makebox(0,0)[cc]{Figure 1}}
\end{picture}

A.Langari and V. Karimipour, \\
A Modified Quantum Renormalization Group for {\bf xxz} spin chain.
\end{center}

\newpage
\unitlength=1.00mm
\special{em:linewidth 0.4pt}
\linethickness{0.4pt}
\begin{center}
\begin{picture}(151.00,109.00)
\put(30.00,80.00){\circle*{2.00}}
\put(40.00,80.00){\circle*{2.00}}
\put(50.00,80.00){\circle*{2.00}}
\put(60.00,80.00){\circle*{2.00}}
\put(70.00,80.00){\circle*{2.00}}
\put(80.00,80.00){\circle*{2.00}}
\put(90.00,80.00){\circle*{2.00}}
\put(100.00,80.00){\circle*{2.00}}
\put(110.00,80.00){\circle*{2.00}}
\put(120.00,80.00){\circle*{2.00}}
\put(130.00,80.00){\circle*{2.00}}
\put(38.00,75.00){\framebox(24.00,10.00)[cc]{}}
\put(68.00,75.00){\framebox(24.00,10.00)[cc]{}}
\put(98.00,75.00){\framebox(24.00,10.00)[cc]{}}
\put(60.00,77.00){\makebox(0,0)[cc]{1}}
\put(70.00,77.00){\makebox(0,0)[cc]{2}}
\put(80.00,77.00){\makebox(0,0)[cc]{3}}
\put(90.00,77.00){\makebox(0,0)[cc]{4}}
\put(100.00,77.00){\makebox(0,0)[cc]{5}}
\put(80.00,99.00){\makebox(0,0)[cc]{Super Block}}
\put(80.00,88.00){\makebox(0,0)[cc]{Block}}
\put(80.00,45.00){\makebox(0,0)[cc]{Figure 2}}
\put(55.00,65.00){\framebox(50.00,30.00)[cc]{}}
\end{picture}

A. Langari and V. Karimipour, \\
A Modified Quantum Renormalization Group for {\bf xxz} spin chain.
\end{center}
\end{document}